\documentclass[prx,aps,onecolumn,showpacs,nofootinbib]{revtex4-2}

\usepackage[dvips]{graphicx} 
\usepackage{amsfonts,amscd,amsmath,amsthm}
\usepackage{enumerate}
\usepackage{epsfig}
\usepackage{subfigure}
\usepackage{diagbox}
\usepackage{xcolor}
\usepackage[colorlinks = true]{hyperref}
\usepackage{physics}
\usepackage{epstopdf}
\usepackage{framed}
\usepackage{multirow}
\usepackage{color}
\usepackage{longtable}
\usepackage{comment}
\usepackage[ruled,vlined]{algorithm2e}
\usepackage[most]{tcolorbox}
\usepackage{longtable}
\usepackage{booktabs}

\graphicspath{{./figure/}}

\usepackage{tikz}
\usetikzlibrary{tikzmark, calc, fit, positioning}
\usetikzlibrary{shapes}
\usetikzlibrary{quantikz}

\newtheorem{proposition}{Proposition}

\newtcolorbox[auto counter]{mybox}[2][]{
	enhanced,
	breakable,
	colback=blue!5!white,
	colframe=blue!75!black,
	fonttitle=\bfseries,
	title=Box \thetcbcounter: #2,#1
}

\begin{document}

\title{Randomness quantification in spontaneous emission}

\author{Chenxu Li}
\thanks{These authors contribute equally to this work.}
\affiliation{Center for Quantum Information, Institute for Interdisciplinary Information Sciences, Tsinghua University, Beijing 100084, China}
\author{Shengfan Liu}
\thanks{These authors contribute equally to this work.}
\affiliation{Center for Quantum Information, Institute for Interdisciplinary Information Sciences, Tsinghua University, Beijing 100084, China}
\author{Xiongfeng Ma}
\email{xma@tsinghua.edu.cn}
\affiliation{Center for Quantum Information, Institute for Interdisciplinary Information Sciences, Tsinghua University, Beijing 100084, China}

\begin{abstract}
 Quantum coherence serves as a fundamental resource for generating intrinsic randomness, yet the quantification of randomness in quantum random number generators (QRNGs) based on spontaneous emission has remained largely phenomenological. Existing randomness analysis lacks rigorous adversarial models and a clear characterization of the role of quantum coherence in these systems. In this work, we develop a comprehensive quantum information-theoretic framework for randomness generation in spontaneous emission processes. We characterize two distinct eavesdropping strategies: one where the adversary directly accesses the atom ensemble, and the other where the adversary accesses only its purification. Our analysis reveals that when randomness is generated through single-photon detection and temporal mode measurements, the QRNG is vulnerable to the first adversary scenario, though it still guarantees a lower bound on intrinsic randomness against the second adversary scenario even under maximal information leakage from the atoms. In contrast, QRNGs based on spatial mode detection and phase fluctuations demonstrate security against both types of adversaries, providing robust randomness generation. Furthermore, we provide a quantitative calculation of intrinsic randomness for these spontaneous-emission-based QRNG schemes.

\end{abstract}

\maketitle

\section{Introduction}
Random numbers play crucial roles across many fields, including numerical simulation~\cite{Metropolis_1949}, cryptography~\cite{Sunar_2009}, and lottery systems. In cryptography, random numbers must exhibit not only statistical uniformity but also security against adversarial prediction. Mainstream random number generators face fundamental security limitations: pseudo-random number generators~\cite{vonN_1951} employ deterministic algorithms whose outputs become predictable given sufficient sequence length, while physical random number generators based on classical physics remain vulnerable to adversaries with side information and computational resources due to the deterministic nature of classical physical laws.

Quantum mechanics provide the possibility to genuine randomness generation through quantum random number generators (QRNGs). According to Born's rule, measuring a superposed state such as $(\ket{0}+\ket{1})/\sqrt{2}$ produces fundamentally unpredictable outcomes. Crucially, while identical measurement statistics can be obtained from classical mixtures like $(\ketbra{0}{0}+\ketbra{1}{1})/2$, which could be generated by classical pseudo-random algorithms, such states cannot guarantee intrinsic randomness. If the measured system is entangled with an external environment, an adversary controlling that environment could perfectly predict the measurement outcomes. Thus, classical mixtures yield no intrinsic randomness despite potentially passing statistical tests.

To quantitatively analyze the origin of randomness, we employ the resource theory of coherence~\cite{Baumgratz_2014,Streltsov_2017}, which quantifies quantum superposition. The relative entropy of coherence—a coherence monotone—has been shown to quantify the amount of intrinsic randomness~\cite{Yuan_2015,Yuan_2019} in terms of its unpredictability to adversaries. This quantification enables discrimination between quantum and classical noise components in measured signals, which is essential for proper post-processing techniques like randomness extraction~\cite{Ma_2013}.

Various QRNG architectures have been developed~\cite{Ma_2016,Herrero_2017,Mannalatha_2023}, with laser-based schemes being particularly prominent due to their high speed and practical implementability. These schemes employ regular lasers~\cite{Xu_2012,Nie_2015} or even LEDs~\cite{Wei_2017,Argillander_2022}. Early approaches generate randomness from photon arrival times~\cite{Stipcevic_2007,Dynes_2008,Nie_2014,Wang_2015,Yan_2015} or spatial positions~\cite{Stefanov_1999,Jennewein_2000,Yan_2014}, though the randomness generation speed of these methods is limited by low single-photon detection rates. Subsequent protocols improved efficiency through coherent detection methods such as homodyne detection~\cite{Gabriel_2010,Jofre_2011,Bruynsteen_2023} and self-heterodyne detection~\cite{Qi_2010,Shen_2010,Xu_2012,Avesani_2021} for randomness generation. Across all these approaches, spontaneous emission serves as the fundamental microscopic origin of intrinsic randomness in laser-based QRNG schemes, particularly highlighted as the origin of random phase fluctuations in lasing fields~\cite{Zhou_2015}.

However, despite experimental maturity, laser-based QRNGs lack a first-principles physical model with proper quantum information-theoretic randomness quantification. For instance, for phase-fluctuation-based QRNGs~\cite{Ma_2013,Zhou_2015}, randomness estimation typically assumes quantum phase fluctuations scale inversely with laser power and approximate Gaussian white noise, arrival-time-based methods~\cite{Dynes_2008,Nie_2014} assume Poissonian photon statistics as their starting point. The absence of rigorous analysis also prevents proper adversarial modeling, compromising information-theoretic security. Even when phenomenological models match observed noise, side-channel vulnerabilities persist, for instance, an eavesdropper with access to the atom ensemble underlying spontaneous emission could exploit atom--field entanglement to predict generated random numbers.  Consequently, establishing a rigorous physical model for randomness generation and security validation in spontaneous emission is essential for proper characterization and security assurance of these QRNGs.

To address this gap, we develop a first-principles analysis of intrinsic randomness in spontaneous-emission-based QRNGs from the rigorous perspective of quantum information theory. Our approach not only clarifies the fundamental origin of randomness but also refines security assumptions across different schemes. Spontaneous emission arises from atom-field interaction, yet detectors only access the resulting radiation field to generate randomness. This raises a critical security question: can QRNGs remain secure if an eavesdropper gains access to the atom ensemble generating the laser field? Surprisingly, we find the answer depends not only on the adversary's capabilities but also on the specific optical property utilized for randomness generation. Our results also quantify the coherence in spontaneous emission via a quantum information-theoretic treatment.

Table~\ref{tab:security} summarizes our main results, quantitatively characterizing the intrinsic randomness in terms of extractible random bits for various detection schemes against two distinct adversary models.


\begin{table}[htbp]
	\centering
	\caption{Intrinsic randomness of different QRNG models against two adversary scenarios. Adversary I: The adversary Eve has direct (passive) access to the atom but cannot manipulate the optical environment. Adversary II: Eve holds a purification (collected earlier emissions/ancilla) of the atomic system but cannot access it directly.}
	\label{tab:security}
	\begin{tabular}{lcc}
		\toprule
		Detection Type & Adversary I & Adversary II \\
		\midrule
		Single-photon & 0 & Eq.~\eqref{RandomnessSinglePhotonLowerBound} \\
		Temporal (arrival time) & 0 & Eq.~\eqref{tempRandomnessLowerBound} \\
		Spatial & Eq.~\eqref{eq:randomnessspatial} & Eq.~\eqref{eq:randomnessspatial} \\
		Phase fluctuation & Eq.~\eqref{eq:randomnessphasefluc} & Eq.~\eqref{eq:randomnessphasefluc} \\
		\bottomrule
	\end{tabular}
\end{table}

The remainder of this paper is organized as follows. Section~\ref{Preliminaries} reviews the Wigner-Weisskopf theory of spontaneous emission and the concept of intrinsic randomness for projective value measurements (PVM) and generalized measurements, establishing the physical and information-theoretic foundations of our work. Section~\ref{Model} introduces our adversary model for randomness in spontaneous emission, classifying eavesdropping strategies into two types based on accessibility to the atom ensemble. Section~\ref{Quantification} applies this adversary model to specific detection schemes, corresponding to different POVM measurements on the radiation system and quantifies the intrinsic randomness.

\section{Preliminaries}
\label{Preliminaries}
In this section, we review the atom--field interaction model in the theory of spontaneous emission and the definition of intrinsic randomness in quantum cryptography.

\subsection{Atom--field interaction}
For an atomic system with two distinct energy levels, the interaction Hamiltonian between it with the radiation field is \cite{Scully_1997}
\begin{equation}\label{totalhamiltonian}
	H=\sum_{\mathbf{k}}\hbar\omega_{\mathbf{k}}a_{\mathbf{k}}^{\dagger}a_{\mathbf{k}}+\frac{1}{2}\hbar\omega\sigma_z+\hbar\sum_{\mathbf{k}}\vec{g}_{\mathbf{k}}(\sigma_+a_{\mathbf{k}}+\sigma_-a^{\dagger}).
\end{equation}

Under the rotating-wave and Markov approximations, following the standard derivation from Wigner-Weisskopf theory \cite{Scully_1997}, we find that a two-level atom interacting with vacuum field evolves as
\begin{equation}\label{wwstate}
	\ket{\psi(t)}=e^{-\Gamma t/2}\ket{e,0}+\sum_{\mathbf{k}}c_k(t)\ket{g,1_{\mathbf{k}}},
\end{equation}
where
\begin{equation}\label{WW}
	c_k(t)=g_{\mathbf{k}}e^{-i\mathbf{k\cdot r}_0}\left[\frac{1-e^{i(\omega-\nu_k)t-\Gamma t/2}}{(\nu_k-\omega)+i\Gamma/2}\right],
\end{equation}
and $\ket{e},\ket{g}$ are the excited state and ground state of the atom, respectively.

In the infinite-time limit, the atom decays to the ground state and the emitted field becomes a superposition over all modes, which has the form
\begin{equation}
	\ket{\psi(\infty)}=\ket{g}\sum_{\mathbf{k}} c_{\mathbf{k}}\ket{1_{\mathbf{k}}}.
\end{equation}

Eq.~\eqref{wwstate} expresses a pure entangled state between the atom and the radiation field, which also has coherence under the measurement basis of different modes, giving rise to intrinsic randomness in the measurement results.

\subsection{Intrinsic randomness}
To analyze the security of randomness generated from a quantum measurement, the user Alice needs to formulate an adversarial scenario, where the adversary Eve may have a certain correlation with Alice's system and try to guess the outcomes of her random numbers. The entropy of outcomes can then be divided into two parts: extrinsic randomness which Eve might know, and intrinsic randomness that Eve has no information about.

Formally, intrinsic randomness is characterized by conditional entropies, which characterize the ignorance of the eavesdropper when they try to predict the outcome of the QRNG. Here we first deal with the case where the measurement is a PVM. Consider an arbitrary state $\rho_A$. after a projective measurement $P$, $\rho_A$ will be dephased to $\rho_A^{P,\mathrm{diag}}$. In the worst case, the adversary $E$ has access to the most side information of the measurement outcomes by holding the purification of $\rho_A$. The joint state before and after the measurement is denoted as $\rho_{AE}$ and $\rho_{A'E}$. For simplicity, in this work we consider the scenario where Alice inputs the same state independently for many rounds of independently and identically performed measurements, i.e. the i.i.d. limit. Then the randomness of a state $\rho$ with respect to a PVM $P$ is characterized by the von Neumann conditional entropy \cite{Yuan_2015,Yuan_2019}
\begin{equation}
	R(\rho,P)=S(A'|E)=S(\rho^{P,\mathrm{diag}})-S(\rho),
\end{equation}
where $S$ is the von Neumann entropy defined as $S(\rho)=-\tr(\rho\log\rho)$. In this work we express entropies in the unit of bits, therefore all logarithmic functions are base 2 unless explicitly stated. $R(\rho,P)$ is a coherence monotone called the relative entropy of coherence, for a resource theory of coherence with respect to the PVM $P$.

A projection measurement is an idealized model for quantum measurements where the user obtains all the information from the detection devices. General measurements are described by positive-operator-valued measures (POVMs). We follow the framework for intrinsic randomness introduced in \cite{Dai_2022}. 

Alice characterizes the source of a QRNG to be in a state $\rho_A$ and performs a POVM measurement $M$. According to the Naimark extension, by introducing an ancillary system $Q$, $M$ can be performed by a PVM $P$ on $AQ$, namely $M_i=\tr_{Q}[P_i(I_A\otimes\sigma_Q)]$. Notice that we use a generalized version of the Naimark extension where both $A$ and $Q$ may be entangled with Eve, while the standard Naimark extension requires $Q$ to be pure. In the worst-case scenario, where Eve is able to gather side information from both $A$ and $Q$, the intrinsic randomness of $\rho$ with respect to a POVM $M$ is defined by \cite{Dai_2022}
\begin{equation}\label{povmrandomness}
	R(\rho,M)=\min_{\sigma,P}R(\rho\otimes\sigma,P),
\end{equation}
where $\{\sigma,P\}$ is a set of Naimark extensions for $M$.

\section{Adversary model for spontaneous emission}
\label{Model}

In this section, we develop the adversary model for randomness in spontaneous emission. We introduce two types of attacks by the adversary that will be distinctly treated in randomness quantification. 

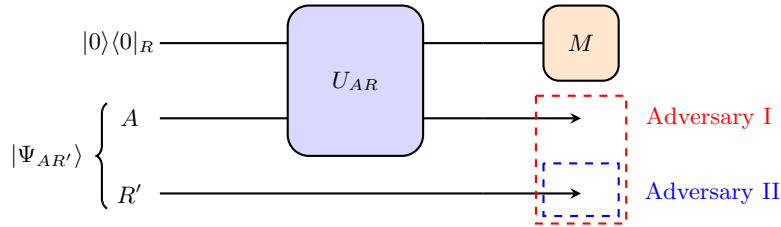
\begin{figure}[htbp]
    \centering
    \begin{tikzpicture}[>=stealth, thick]

    \tikzstyle{unitary} = [draw, fill=blue!15, rounded corners=8pt, minimum height=2cm, minimum width=1.8cm]
    \tikzstyle{measure} = [draw, fill=orange!20, rounded corners=6pt, minimum height=1cm, minimum width=1cm]
    
    \node at (-2.35,1.0) {$|0\rangle\langle0|_R$};
    \node at (-2.2,0) {$A$};
    \node at (-2.2,-1.0) {$R'$};
    
    \draw [decorate, decoration={brace, amplitude=4pt}] (-2.5,-1.2) -- (-2.5,0.2);
    \node at (-3.3,-0.5) {$\ket{\Psi_{AR'}}$};
    
    \draw (-1.8,1.0) -- (2.5,1.0);
    \draw (-1.8,0) -- (2.5,0);
    \draw (-1.8,-1.0) -- (2.5,-1.0);
    
    \node[unitary] at (0.8,0.5) {$U_{AR}$};
    
    \draw (2.5,1.0) -- (3.8,1.0);
    \draw [->](2.5,0) -- (3.8,0);
    \draw [->](2.5,-1.0) -- (3.8,-1.0);
    
    \node[measure] at (3.8,1.0) {$M$};

    \draw[red, dashed, thick] (3.2,0.3) rectangle (4.4,-1.4);
    \draw[blue, dashed, thick] (3.3,-0.6) rectangle (4.3,-1.3);
    
    \node[red] at (5.5,0) {Adversary I};
    \node[blue] at (5.55,-1) {Adversary II};
    
    \end{tikzpicture}
    \caption{Schematic illustration of the adversary scenario. We consider the systems $A$, $R$, and $R'$, which correspond to the atom, the emitted radiation, and the purification of the initial atomic system. The input state for $R$ is the vacuum state $\ketbra{0}{0}_R$ with no photons. The spontaneous emission process is described as a unitary evolution $U_{AR}$. The two types of adversary have access to $AR'$ and $R'$, respectively. The user performs a POVM $M$ on system $R$ and the measurement outcome is used to generate random numbers. Different implementations of $M$ correspond to different QRNG schemes discussed in this work.}
    \label{fig:Adversary}
\end{figure}

As shown in Fig.~\ref{fig:Adversary}, we denote the atomic system by $A$ and the emitted radiation by $R$. The spontaneous emission process is described by a unitary evolution $U_{AR}$. Alice collects the photons and obtains randomness from certain optical properties of the photons, which correspond to specific POVM measurements on $R$. Recall the example of an attack on the state $(\ketbra{0}+\ketbra{1})/2$: such an attack by an adversary can be described by a measurement on the “environment” degrees of freedom. We thus categorize adversaries into two types, based on their ability to access different degrees of freedom:
\begin{itemize}
    \item Adversary I: The adversary has direct access to the atom ensemble, meaning that she can manipulate the state of the atom and perform her own measurement on the atom at will, but cannot manipulate the optical environment (cannot inject light, alter cavity--vacuum coupling, or perform intercept--resend attacks on the channel).
    \item Adversary II: The state of the atom ensemble is hidden from the adversary. The adversary can at best obtain information from side information to perform her side attacks, for example, she can collect all the information of the previously emitted photons. The ancillary system in the purification is denoted as system $R'$. The atom is initially purified as the state $\ket{\Psi_{AR'}}$.
\end{itemize}
In both cases, system $A$ is not accessible to Alice.

For example, if we take the state $\ket{\Psi_{AR'}}$ as the state in Eq.~\eqref{wwstate} when $t=t_0$, the joint input state is
\begin{equation}
	\ket{\Psi_{in}}_{AR'R}=\left(e^{-\Gamma t_0/2}\ket{e}_A\ket{0}_{R'}+\sum_{\mathbf{k}}c_{\mathbf{k}}(t_0)\ket{g}_A\ket{1_{\mathbf{k}}}_{R'}\right)\otimes\ket{0}_R.
\end{equation}
The spontaneous emission by the unitary $U_{AR}$ acts nontrivially only on excited atom states:
\begin{equation}
	U_{AR}:\begin{cases}
		\ket{e}_A\ket{0}_R\mapsto e^{-\Gamma t/2}\ket{e}_A\ket{0}_R+\sum_{q}c_{q}(t)\ket{g}_A\ket{1_{q}}_R\\
		\ket{g}_A\ket{0}_R\mapsto \ket{g}_A\ket{0}_R
	\end{cases},
\end{equation}
where $q$ denotes the modes of Alice's field. Thus the output state for the total system is 
\begin{equation}	
	\begin{aligned}\label{eq:wwstate_Eve}
		\ket{\Psi_{out}}_{AR'R}&=(U_{AR}\otimes I_{R'})\ket{\Psi_{in}}_{AR'R}\\
		&=e^{-\Gamma t_0/2}\left(e^{-\Gamma t/2}\ket{e}_A\ket{0}_R+\sum_{q}c_{q}(t)\ket{g}_A\ket{1_{q}}_R\right)\ket{0}_{R'}\\
		&\quad +\ket{g}_A\ket{0}_R\otimes\sum_{\mathbf{k}}c_{\mathbf{k}}(t_0)\ket{1_{\mathbf{k}}}_{R'}.
	\end{aligned}
\end{equation}
To perform randomness quantification, we need to answer the question of how much intrinsic randomness exists in Eq.~\eqref{eq:wwstate_Eve} for a certain POVM measurement on system $R$, given a certain adversary model. In the most general case,  

We emphasize that in this work we remain in the trusted-device scenario, in contrast to the developing field of (semi)-device-independent QRNGs \cite{Pironio_2010,Chaturvedi_2015,Cao_2015,Liu_2018,Michel_2019,lin_2023,Wu_2023,Nie_2024,Bertapelle_2025}, where trust or characterization for certain devices is removed. For example, we do not consider detection side channels targeting detector imperfections, or the possibility of an intercept--resend attack targeting the quantum source.

\section{Randomness Quantification in Spontaneous-Emission-Based QRNGs}
\label{Quantification}
In this section, we try to quantify the intrinsic randomness of the spontaneous emission state with respect to different measurement schemes.

\subsection{Single-photon detection QRNG}
We first consider single-photon-detection-based QRNGs. We model this type of QRNG by using one single photon detector to capture the emitted photons. Randomness is generated from the signal of detecting or not detecting a photon in a given period of time. The POVM conducted on system $R$ is the PVM $\{P_i\}=\{\ketbra{0}{0}_R,\sum_{\mathbf{k}}\ketbra{\mathbf{k}}{\mathbf{k}}_R\}$ that coarse-grains the spatial degree of freedom. The outcome $i=0,1$ corresponds to the event of not detecting a photon and detecting one, then the state space of photons can be simplified to a qubit Hilbert space. QRNGs that involves measuring spatial degree of freedom will be discussed in later subsections.

The example in Eq.~\eqref{eq:wwstate_Eve} can also be simplified to 
\begin{equation}
    \ket{\Psi_{out}}_{AR'R}
    = e^{-\Gamma t_0/2}\left(e^{-\Gamma t/2}\ket{e}_A\ket{0}_R
    + e^{i\theta}\sqrt{1-e^{-\Gamma t}}\ket{g}_A\ket{1}_R\right)\otimes\ket{0}_{R'}
		+ e^{i\theta}\sqrt{1-e^{-\Gamma t_0}}\ket{g}_A\ket{0}_R\otimes\ket{1}_{R'}
\end{equation}
for a certain phase factor $\theta$.

By tracing out system $A$ and $R'$ in Eq.~\eqref{eq:wwstate_Eve}, the state of system $R$ becomes
\begin{equation}
\rho_R = \begin{pmatrix}
1 - e^{-\Gamma t_0}\left(1 - e^{-\Gamma t}\right) & 0 \\
0 & e^{-\Gamma t_0}\left(1 - e^{-\Gamma t}\right)
\end{pmatrix}.
\end{equation}
This is a classical state containing no intrinsic randomness when we consider the type I adversary model in which Eve takes control of the atom. In this case, the state $\ket{\Psi_{out}}_{AR'R}$ is the purification of $\rho_R$ and Eve can predict the measurement outcomes on $R$ by performing her own measurements on system $AR'$.

For type II adversary who does not have direct access to $A$, notice that 
\begin{equation}
    \Pr(i)=\tr[(P_i\otimes I_A)U^\dagger_{AR}(\rho_A\otimes\ketbra{0}{0}_R)U_{AR}],
\end{equation}
then we equivalently seek the value of
\begin{equation}
    R(\rho_A\otimes\ketbra{0}{0}_R,\Pi),
\end{equation}
where $\Pi$ is a PVM defined by $\Pi_i=U_{AR}(P_i\otimes I_A)U^\dagger_{AR}$, which is a standard Naimark extension of a POVM measurement on the initial state of $A$ before the spontaneous emission. Compared to Eq.~\eqref{povmrandomness}, we do not require the minimization over all Naimark extensions because the physical model allows for only one specific Naimark extension. Also, the system $R$, which is now treated as the ancillary system, is intially in a pure state $\ketbra{0}_R$, so that we do not need to worry it being entangled with the eavesdropper. 

The equivalent POVM has two elements
\begin{equation}
	E_0=\begin{pmatrix}
		1&0\\0&e^{-\Gamma t}
	\end{pmatrix},\quad E_{1}=\begin{pmatrix}
		0&0\\0&1-e^{-\Gamma t}
	\end{pmatrix}
\end{equation}
with corresponding Kraus operators
\begin{equation}
    M_0=\begin{pmatrix}
        1 & 0\\
        0 & e^{-\Gamma t/2}
    \end{pmatrix},\quad 
    M_{1}=\begin{pmatrix}
        0 & \sqrt{1-e^{-\Gamma t}}\\
        0 & 0
    \end{pmatrix}.
\end{equation}
After the POVM measurement that includes the joint effect of a Hamiltonian evolution and photon measurement process, the post-measurement state becomes $M_i \rho_A M^\dagger_i/\tr(\rho_A E_i)$ for $i=0,1$. Taking system $R$ as the ancillary system in the Naimark extension, the joint post-measurement state are given by
\begin{equation}
\begin{gathered}
\tau^0_{AR}=\frac{1}{\rho_{00}+e^{-\Gamma t} \rho_{11}}\left(\begin{array}{cc}
\rho_{00} & \sqrt{e^{-\Gamma t}} \rho_{01}\\
\sqrt{e^{-\Gamma t}} \rho_{10} & e^{-\Gamma t} \rho_{11}
\end{array}\right)_A \otimes\ketbra{0}_R,\\ \tau^1_{AR}=\begin{pmatrix}
    1 & 0\\ 0 & 0
\end{pmatrix}_A\otimes\ketbra{1}_R
\end{gathered}
\end{equation}
with probability $
\operatorname{Pr}(0)=\rho_{00}+e^{-\Gamma t} \rho_{11},\operatorname{Pr}(1)=\rho_{11}\left(1-e^{-\Gamma t}\right)$. Here $\rho_{00},\rho_{01},\rho_{10},\rho_{11}$ are the entries of the density matrix $\rho_A$. 

The dephased state is of a block diagonal form in the joint Hilbert space of $AR$, therefore, the intrinsic randomness can be calculated as
\begin{equation}
\label{eq:randomnesssinglephoton}
\begin{aligned}
R(\rho_A\otimes \sigma_R,\Pi)&=-S(\rho_A)+S\left[\operatorname{Pr}(0)\tau^0_{AR}+\operatorname{Pr(1)\tau^1_{AR}}\right]\\
&= -S(\rho_A)-\rho_{11}\left(1-e^{-\Gamma t}\right) \log \left[\rho_{11}\left(1-e^{-\Gamma t}\right)\right] -\mu_1(t) \log \mu_1(t)-\mu_2(t) \log \mu_2(t),
\end{aligned}
\end{equation}
where \begin{equation}
    \mu_{1,2}(t)=\frac{1}{2}\left[\rho_{00}+e^{-\Gamma t} \rho_{11} \pm \sqrt{\left(\rho_{00}-e^{-\Gamma t} \rho_{11}\right)^2+4 e^{-\Gamma t}\left|\rho_{01}\right|^2}\right].
\end{equation}

Eve's most effective attack is to gather all the side information of the atomic system, for instance, collect all the previous emissions. Under this type of attack, $|\rho_{01}|$ vanishes and $\rho_A$ becomes incoherent. Nevertheless, in this case we can still generate randomness that is unpredictable by Eve. The reason is that our equivalent POVM model acts on the atomic system before the emission process $U_{AR}$, and the unitary evolution $U_{AR}$ generates fresh coherence that can be harvested by Alice, even if it acts on an incoherent state.

Eq.~\eqref{eq:randomnesssinglephoton} gives the most general form of randomness for single-photon measurement scheme and depends on a full density matrix of $A$, which has off-diagonal terms that are hard to characterize in experiments when Alice do not have access of the atom. We utilize the following proposition to find a lower bound of Eq.~\eqref{eq:randomnesssinglephoton}:

\begin{proposition}
    For fixed diagonal entries $\rho_{11}$ and $\rho_{00}=1-\rho_{11}$, the function $R(\rho_A)$ defined in Eq.~\eqref{eq:randomnesssinglephoton} is strictly increasing with respect to $|\rho_{01}|$.
\label{proposition1}
\end{proposition}

The proof of Proposition \ref{proposition1} can be found in Appendix \ref{ProofProposition1}. Therefore, the lower bound of the randomness can be obtained by setting $|\rho_{01}|=0$, then we have the simplification $\mu_1=\rho_{00}$ and $\mu_2=e^{-\Gamma t}\rho_{11}$, and thus:
\begin{equation}
\label{RandomnessSinglePhotonLowerBound}
    R\ge\rho_{11}\left[-(1-e^{-\Gamma t}) \log (1-e^{-\Gamma t})-e^{-\Gamma t} \log e^{-\Gamma t}\right].
\end{equation}

The lower bound only involves the $\rho_{11}$ element, i.e., the population of the atom, which can be easily characterized by measuring the photon emission rate $I=\Gamma \rho_{11}$. Another possible method is to couple the atom ensemble with a heat bath and initialize its state as a Gibbs state, $\rho_{11}$ can be obtained from the temperature of the heat bath. 

\subsection{Temporal mode QRNG}
Another measurement scheme is the temporal mode measurement, which records the arrival time of the detected photon. Since we can divide the arrival time into multiple time bins, the advantage of this scheme is that it can generate multiple random bits from every detection event \cite{Stipcevic_2007,Dynes_2008,Nie_2014, Khanmohammadi_2015}. 

We introduce $n$ time bins for the total time interval $[0,t]$. For a emitted photon that is not detected, the QRNG outputs 0. If it is detected and the arrival time of the photon falls into one of the time bins, the QRNG outputs the number of that time bin. Therefore for every photon emitted, a random number with $n+1$ possible values can be generated. Compared to single photon detection that only outputs a binary value, the entropy source has a higher dimension which correspond to more extractable random bits. Some temporal mode based approaches \cite{Stipcevic_2007} measure time interval between two detection events and randomness is generated from the fluctuation of the quantity. Nevertheless, we can treat them as an extra postprocessing method on the entropy source of our QRNG model, therefore not affecting any generality. 

To simplify notations, we use the amplitude damping channel to describe the spontaneous emission process. By using the channel description, we implicitly use the same Markovian approximation in Weisskopf-Wigner theory. Suppose every time bin has the same length, then in each time bin the system undergoes a quantum channel described by
\begin{equation}
\label{eq:AmplitudeDamping}
	\begin{aligned}
		\ket{0}_A\ket{0}_R\mapsto&\ket{0}_{A}\ket{0}_R,\\
		\ket{1}_A\ket{0}_R\mapsto&\sqrt{1-p}\ket{1}_A\ket{0}_R+\sqrt{p}\ket{0}_A\ket{1}_R.
	\end{aligned}
\end{equation}
with $p=1-e^{-\Gamma t/n}$. Here we also neglect the spatial degree of freedom for photons. The measurement done in temporal mode can be understood as follows: in each time interval, the system undergoes the amplitude damping channel, and then a PVM is performed on it. For simplicity, we have neglected device imperfections such as detector dead time and dark counts. In this setting, photons in system $R$ that belong to different time bin are orthogonal and can be perfectly distinguishable by a PVM. Therefore, without loss of generality, we treat system $R$ as a Naimark extension with dimension $2^n$, which has the state space as a tensor product of $n$ qubit Hilbert spaces. 

Similar to the treatment of single-photon detection, the temporal mode measurement on $R$ is modeled as a POVM on system $A$:
\begin{equation}
	    E_0=\begin{pmatrix}
    		1&0\\0&(1-p)^n
    	\end{pmatrix}, \quad
        E_{k\ge 1}=\begin{pmatrix}
    		0&0\\0&p(1-p)^{k-1}
    	\end{pmatrix}.
\end{equation}
The corresponding Kraus operators are
\begin{equation}
        M_0=\begin{pmatrix}
            1 & 0\\
            0 & \sqrt{(1-p)^n}
        \end{pmatrix}, \quad
        M_{k\ge 1}=\begin{pmatrix}
            0 & \sqrt{p(1-p)^{k-1}}\\
            0 & 0
        \end{pmatrix}.
\end{equation}

The probabilities to get measurement output $0$ to $n$ are
\begin{equation}	
	\begin{gathered}	
		\operatorname{Pr}(0)=\tr(\rho_AE_0)=1-\rho_{11}+(1-p)^n\rho_{11},\\
		\operatorname{Pr}(k\geq 1)=\tr(\rho_AE_{k\geq 1})=(1-p)^{k-1}p\rho_{11}.	
	\end{gathered}	
\end{equation}
With the corresponding post measurement states being
\begin{equation}
	\begin{gathered}
		\tau_{AR}^0=\frac{1}{\rho_{00}+\rho_{11}(1-p)^n}\begin{pmatrix}
			\rho_{00}&\sqrt{(1-p)^n}\rho_{01}\\\sqrt{(1-p)^n}\rho_{10}&(1-p)^n\rho_{11}
		\end{pmatrix}_A\otimes\bigotimes_{i=1}^{n}\ketbra{0}_{R_i},\\
		 \tau_{AR}^{k\ge 1}=\frac{1}{\rho_{11}(1-p)^{k-1}p}\begin{pmatrix}
			(1-p)^{k-1}p\rho_{11}&0\\0&0
		\end{pmatrix}_A\otimes\ketbra{1}_{R_k}\otimes\bigotimes_{i\neq k}^{n}\ketbra{0}_{R_i}.
	\end{gathered}	
\end{equation}
Since the supports of these states are orthogonal, we can obtain the intrinsic randomness here by 
\begin{equation}
\label{tempRandomness}
	R(\rho_A\otimes \sigma_R,\Pi)=-S(\rho_A)-\sum_{k=1}^{n}(1-p)^{k-1}p\rho_{11}\log[(1-p)^{k-1}p\rho_{11}]-\mu_1\log\mu_1-\mu_2\log\mu_2.
\end{equation}
where
\begin{equation}
	\mu_{1,2}=\frac{1}{2}\left[\rho_{00}+(1-p)^n\rho_{11}\pm\sqrt{(\rho_{00}-(1-p)^n\rho_{11})^2+4(1-p)^n|\rho_{01}|^2}\right].
\end{equation}
The previous result of single-photon detection can be regarded as a special case of temporal mode measurement where there is only one time bin. Therefore, the quantification given by Eq.~\eqref{tempRandomness} also belongs to the setting where Eve can only hold at most the purification of the atom state, instead of directly accessing it. It is also straightforward from Eq.~\eqref{eq:wwstate_Eve} to see that the detection scheme is also insecure when Eve has access to the atom, because by monitoring the atom state in each time bin by measuring on system $A$ and $R'$, Eve is able to learn the state of $R$ in every time bin and thus predict Alice's random numbers.

From Eq.~\eqref{tempRandomness} we can see that the off-diagonal terms of $\rho_A$ affect the intrinsic randomness. By using an argument similar to Proposition \ref{proposition1}, we can also show that Eq.~\eqref{tempRandomness} is strictly increasing with respect to the off-diagonal term of $\rho_A$. Therefore, we can find the lower bound of Eq.~\eqref{tempRandomness} by setting $|\rho_{01}|=0$:
\begin{equation}
    \begin{aligned}
R\ge & -S\left(\rho_A\right)-\sum_{k=1}^n(1-p)^{k-1} p \rho_{11} \log \left[(1-p)^{k-1} p \rho_{11}\right]-\rho_{00} \log \rho_{00}-(1-p)^n \rho_{11} \log \left[(1-p)^n \rho_{11}\right] \\
= & \rho_{00} \log \rho_{00}+\rho_{11} \log \rho_{11}-\sum_{k=1}^n(1-p)^{k-1} p \rho_{11}\left[\log \rho_{11}+\log p+(k-1) \log (1-p)\right] \\&-\rho_{00} \log \rho_{00}-(1-p)^n \rho_{11}\left[n \log (1-p)+\log \rho_{11}\right] .
\end{aligned}
\end{equation}
Using the equality $\sum_{k=1}^n(k-1)x^{k-1}=[x-nx^n+(n-1)x^{n+1}]/(1-x)^2$, we can further simplify the result as
\begin{equation}
\label{tempRandomnessLowerBound}
\begin{aligned}
    R&\ge \rho_{11}\frac{1-(1-p)^n}{p}\left[-p\log p-(1-p)\log(1-p)\right]\\
    &=\rho_{11}\frac{1-e^{-\Gamma t}}{1-e^{-\Gamma t/n}}\left[-(1-e^{-\Gamma t/n})\log (1-e^{-\Gamma t/n})-e^{-\Gamma t/n}\log e^{-\Gamma t/n}\right].
\end{aligned}
\end{equation}
Similar to discussions in the previous subsection, $\rho_{11}$ can be assessed from the photon emission rate.  Eq.~\eqref{tempRandomnessLowerBound} can be treated as a generalization of Eq.~\eqref{RandomnessSinglePhotonLowerBound}, since single-photon detection can be viewed as a arrival time measurement where there is only one time bin available.

\begin{figure}[htbp]
    \centering
    \includegraphics[width=0.7\linewidth]{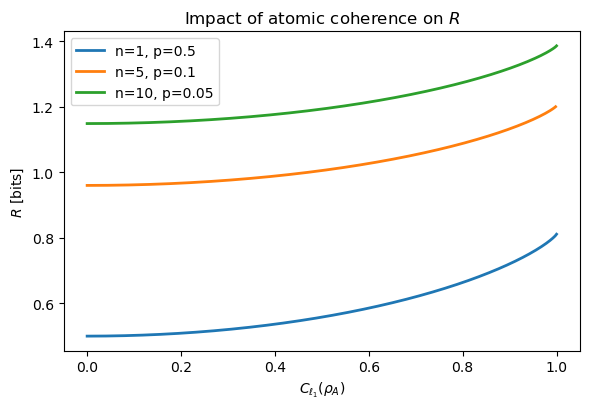}
    \caption{Impact of atomic coherence on intrinsic randomness, for temporal measurement and an adversary with no access to the atomic system. The curves are plotted with different numbers of time bin and coherence is quantified using the $l_1$ norm of coherence. When $n=1$, temporal measurement becomes the single-photon detection. The diagonal terms of the state $\rho_A$ is set to be both 1/2 and its possible $l_1$ norm of coherence ranges from 0 to 1. The randomness is expressed in terms of number of extractable bits.}
    \label{fig:tempRandomness}
\end{figure}

Now we briefly remark on the physical interpretation of the result. The sum of the absolute values of the off-diagonal terms $\sum_{i\neq j}|\rho_{ij}|$ is a coherence monotone called the $l_1$ norm of coherence \cite{Baumgratz_2014}. Operationally, for $\rho_A$ with the same diagonal terms but with different amounts of coherence, it quantifies the leakage of side information to the adversary holding its purification. In Fig.~\ref{fig:tempRandomness}, we numerically investigate the impact of atomic coherence on the intrinsic randomness generated from temporal measurements, which characterizes the influence of information leakage from the atom to the adversary. Our analytical and numerical results both demonstrate that even when there is no coherence from the atom, indicating maximal information leakage, there still remains intrinsic randomness, which serves as a lower bound of extractable randomness from the collected noise signal. Randomness can also be increased by simply adding more time bins, as long as detector inefficiencies can be neglected.

We also remark is that in the continuous-time limit, we have $p=\Gamma\Delta t\ll1$, which indicates
\begin{equation}
	\begin{aligned}
		\operatorname{Pr}(\text{No\ detection})&=1+(e^{-\Gamma T}-1)\rho_{11},\\
		\operatorname{Pr}(t)dt&=\rho_{11}e^{-\Gamma t}dt.
	\end{aligned}
\end{equation}
Under this limit, the probability of detecting a photon at $t$ obeys an exponential distribution. If we have many i.i.d.
atoms, then we obtain the time-of-arrival statistics, where the number of photons detected within a time period obeys
a Poisson distribution. We thus recover the main assumption in temporal mode QRNGs \cite{Dynes_2008,Nie_2014} from first principles.

\subsection{Spatial mode QRNG}
Spatial mode QRNGs extract randomness from the direction in which a spontaneously emitted photon is detected. 
Operationally, an array of photon detectors is placed at different spatial positions, each corresponding to a distinct optical mode \cite{Yan_2014,Burri_2014}.
The fundamental source of unpredictability is the quantum superposition over radiation modes created by spontaneous emission.

Starting from Eq.~\eqref{eq:wwstate_Eve} and tracing out the atom and adversary, the emitted field state takes the form
\begin{equation}\label{spatialstate}
    \rho_{R}
    = \left(e^{-\Gamma (t+t_0)}+1-e^{-\Gamma t_0}\right)\ketbra{0}+\left(\sum_{\mathbf{k}} c_{\mathbf{k}} \ket{1_{\mathbf{k}}}_R \right)\left(\sum_{\mathbf{k}} c_{\mathbf{k}}^* \bra{1_{\mathbf{k}}}_R \right),
\end{equation}
which is a coherent superposition of single-photon excitations across spatial modes $\mathbf{k}$ with amplitudes $c_{\mathbf{k}}$.

To model a realistic measurement, we partition the optical modes into $m$ disjoint subsets $\{K_i\}$, where $K_i$ corresponds to the set of modes collected by the $i$th detector. 
The measurement is thus described by the projective POVM
\begin{equation}
    \left\{\sum_{\mathbf{k}\in K_1}\ketbra{1_{\mathbf{k}}},\ldots,\sum_{\mathbf{k}\in K_m}\ketbra{1_{\mathbf{k}}}\right\}.
\end{equation}
Defining normalized mode states
\begin{equation}
    \ket{1_{\phi_i}}
    = \frac{\sum_{\mathbf{k}\in K_i} c_{\mathbf{k}} \ket{1_{\mathbf{k}}}}
           {\sqrt{\sum_{\mathbf{k}\in K_i} |c_{\mathbf{k}}|^2}},
\end{equation}
we rewrite Eq.~\eqref{spatialstate} as
\begin{equation}
\rho_{R}=\left(e^{-\Gamma (t+t_0)}+1-e^{-\Gamma t_0}\right)\ketbra{0}+\left(\sum_{i=1}^{m}c_i\ket{1_{\phi_i}}\right)\left(\sum_{i=1}^{m}c_i^*\bra{1_{\phi_i}}\right),
\end{equation}
where $|c_i|^2 = \sum_{\mathbf{k}\in K_i} |c_{\mathbf{k}}|^2$ is the probability that a photon is emitted into detector $i$'s acceptance region, which can be directly obtained from the clicking probability $p_i$ of each detector in the experiment. By writing $p_i=|c_i|^2$, the intrinsic randomness in spatial mode QRNGs can be written as 
\begin{equation}\label{eq:randomnessspatial}
    R = -\sum_{i=1}^{m} p_i \log p_i,
\end{equation}
which corresponds to the Shannon entropy of the spatial emission distribution.

Unlike single-photon or temporal mode QRNGs, the measurement in the spatial basis collapses $\sum_{\mathbf{k}} c_{\mathbf{k}} \ket{1_{\mathbf{k}}} \rightarrow\ket{1_{\phi_i}}$, 
breaking coherence by breaking superposition over spatial radiation modes induced by spontaneous emission. 
Since these directional components arise from vacuum-induced spontaneous emission and not from the atomic internal state,
even an adversary with joint access to $A$ and $R'$ cannot predict the emission direction.
Thus, spatial mode QRNGs generate intrinsic randomness against both adversary models I and II.


\subsection{Quantum phase fluctuation based QRNG}
In addition to spontaneous-emission processes with discrete measurement, randomness can also be extracted from the detection of the phase fluctuations of a laser field. 
A common implementation~\cite{Xu_2012, Zhou_2015, Nie_2015} employs a planar lightwave circuit Mach--Zehnder interferometer (PLC--MZI), 
which interferes two delayed temporal modes with time delay $\tau$ of the laser output. In previous works, the quantity of randomness in phase fluctuation based QRNG has been quantified \cite{Zhou_2015}, in this subsection we show that from our model we can also derive the same result. The problem in expressing phase fluctuation using quantum information in previous works is that the quantum phase fluctuation is directly regarded as white noise \cite{Xu_2012, Zhou_2015}, which needs to be clarified.

From the perspective of quantum information, denote the cavity system as $L$, the intracavity field during each emission interval can be modeled as a coherent state
$\ket{\alpha e^{i\phi_m}}_L$, which is coupled to the vacuum field populated by spontaneous emission $\rho_R$ in Fig.~\ref{fig:Adversary}. 
During each interval, spontaneous emission couples the cavity mode to a continuum of vacuum modes. Following Eq.~\eqref{eq:wwstate_Eve}, the state of each single spontaneous emission event where emitted photon exists lies in a superposition $\sum_{\mathbf{k}}c_{\mathbf{k}}\ket{1_{\mathbf{k}}}_R$, 
where the amplitudes $c_{\mathbf{k}}$ carry random phases determined by vacuum fluctuations. Tracing out the vacuum modes transfers this microscopic mode superposition into a random phase increment $\delta\phi_m=\phi_{m+1}-\phi_m$ of the intracavity field.

Microscopically, the cavity annihilation operator $a_L$ couples to external vacuum modes $\{b_\omega\}_R$ via
\begin{equation}
	U_{\mathrm{int},LR}=\exp\left[-i\int d\omega g(\omega)(a_L^\dagger {b_{\omega}}_R+a_L{b_{\omega}^\dagger}_R)\right],
\end{equation}
where $g(\omega)$ denotes the coupling strength between the intracavity field mode $a_L$ and each vacuum mode ${b_{\omega}}_R$. 
Tracing out the vacuum field coupled to the cavity, populated by spontaneous emission yields that the effective intracavity map for the intracavity state $\rho_L$ can be written as
\begin{equation}\label{eq:dephasing_channel}
	\mathcal{E}_{\mathrm{SE}}(\rho_L)=\mathrm{Tr}_{R}\left[U_{\mathrm{int}}(\rho_L\otimes \ket{0}\bra{0}_R)U_{\mathrm{int}}^\dagger\right]=\int d(\delta\phi)p(\delta\phi)e^{-i\delta\phi \hat{n}}\rho e^{i\delta\phi \hat{n}},
\end{equation}
where $p(\delta\phi)$ is the distribution of phase kicks induced by vacuum fluctuations and $\hat{n}$ is the photon-number operator, and the subsystem $\rho_R$ being traced out represents the external vacuum modes of the field populated by spontaneous emission.

Each random phase increment partially destroys the off-diagonal coherence between photon-number components of the field, resulting in a gradual diffusion of the optical phase.
Taking the average over all independent spontaneous emission events in this emission interval, the accumulated phase evolution obeys the diffusion equation $d\phi/dt=\xi(t)$ with 
\begin{equation}
	\langle\xi(t)\xi(t')\rangle=2D_\phi\delta(t-t'),
\end{equation}
which corresponds to the ``Gaussian white noise'' model used in previous analyses for phase fluctuation based QRNGs~\cite{Xu_2012, Zhou_2015, Nie_2015}, meaning that the source of randomness in the 
phase fluctuation based QRNGs is the superposition over different modes. Therefore, we have recovered the phenomenological model in which the phase evolution obeys Wiener process, which implicitly imposed the assumption that Eve cannot access the environment during the cavity-vacuum interaction. If this assumption fails, the phenomenological model and the security of phase noise based QRNGs will also be undermined.

The following analysis simply follows the existing analysis \cite{Zhou_2015}. Thus, we omit the steps and directly present the quantity of randomness \cite{Zhou_2015} here
\begin{equation}\label{eq:randomnessphasefluc}
    \begin{aligned}
        R=-\log\left[2\Phi\left(\frac{\lambda}{\sqrt{\tau}}\right)-1\right].
    \end{aligned}
\end{equation}
Where $\Phi(x)$ is the cumulative distribution function of a standard Gaussian distribution, and 
\begin{equation}
    \lambda=\frac{a}{4\pi P}\sqrt{\frac{\tau_c}{A}},
\end{equation}
where $a$ denotes the width of the voltage interval, $P$ is the output power of the laser, and $\tau_c$ denotes the coherence time.

Next, we show which coherence is broken in this process. The PLC--MZI interferometer measures the relative phase between two consecutive temporal modes. 
The joint field state of two successive intervals can be expressed as
\begin{equation}\label{eq:two_modes}
	\ket{\Psi_{12}}=\ket{\alpha e^{i\phi_m}}_1\otimes\ket{\alpha e^{i(\phi_m+\delta\phi_m)}}_2.
\end{equation}
The interferometer mixes the two modes on a balanced beam splitter and measures the interference operator
\begin{equation}
	\hat{V}=a_1^{\dagger}a_2+a_1a_2^{\dagger}.
\end{equation}
The expectation value of this operator for the state Eq.~\eqref{eq:two_modes}
\begin{equation}\label{eq:voltage}
	\langle\hat{V}\rangle \propto |\alpha|^2\cos(\delta\phi_m)
\end{equation}
gives the voltage output, where randomness arises from the spontaneous emission-induced $\delta \phi_m$.
The broken coherence here is the microscopic superposition over external field modes produced by spontaneous emission.
Security thus relies only on the assumption that Eve cannot access the environment during the cavity-vacuum interaction; under this assumption, phase-fluctuation QRNGs are intrinsically comparable to spatial-mode QRNGs in security.
Even if an adversary Eve has access to the atom ensemble, Eve cannot obtain the result of the random number generated if the emitted field cannot be accessed. In this 
aspect, quantum phase fluctuation based QRNGs are relatively secure with respect to temporal mode and single-photon detection QRNGs.

\section{Conclusion}

In this work, we have developed a comprehensive quantum information-theoretic framework for analyzing QRNGs based on spontaneous emission. By modeling the detection of spontaneous emission as a coherence breaking process in the joint atom-field system, we precisely identified the physical origin of intrinsic randomness across different QRNG schemes. Our approach provides rigorous quantification for several QRNG protocols and establishes a security analysis framework that accounts for potential attacks targeting the atom ensemble itself.

We demonstrate that single-photon and temporal mode QRNGs rely on the collapse of atom-field superpositions, whereas spatial mode and quantum phase fluctuation QRNGs derive their randomness from spontaneous emission-induced superpositions over the modes of the emitted light. This distinction clarifies the trust hierarchy among these protocols: while some schemes require partial trust in the atomic ensemble, others maintain intrinsic randomness even when the atomic subsystem is accessible to an adversary. Our framework thus provides a unified perspective connecting spontaneous emission, quantum coherence, and randomness generation, serving as a foundation for future analysis of spontaneous-emission-based QRNG protocols and quantum randomness certification.

\section*{Acknowledgement}
This work was supported by the National Natural Science Foundation of China (Grants No. 12174216 and No. 12575023) and the Quantum Science and Technology-National Science and Technology Major Project (Grants No. 2021ZD0300804 and No. 2021ZD0300702).

\appendix

\section{Proof of Proposition \ref{proposition1}}
\label{ProofProposition1}
Let $c:=|\rho_{01}|$ with $\rho_{00}$ fixed, then we have
\begin{equation}
\begin{aligned}
    \frac{d R}{d c}&=\frac{2 c}{\Delta} \log \frac{1+\Delta}{1-\Delta}-\frac{2 e^{-\Gamma t} c}{\Delta'} \log \frac{F+\Delta'}{F-\Delta'}\\
    &=4c\left[\frac{\operatorname{arctanh}(\Delta)}{\Delta}-\frac{e^{-\Gamma t}}{F}\frac{\operatorname{arctanh}(\Delta'/F)}{\Delta'/F}\right],
\end{aligned}
\end{equation}
where $F:=\rho_{00}+e^{-\Gamma t}\rho_{11}$, $\Delta:=\sqrt{(\rho_{11}-\rho_{00})^2+4c^2}$ and $\Delta':=\sqrt{(\rho_{00}-e^{-\Gamma t}\rho_{11})^2+4e^{-\Gamma t}c^2}$. 

By using the identity
\begin{equation}
    \frac{\operatorname{arctanh}(x)}{x}=\int_0^1\frac{ds}{1-x^2s^2}\qquad (0<x<1),
\end{equation}
we have the integral representation
\begin{equation}
\label{integral}
    \frac{dR}{dc}=4c\int_0^1 \left[\frac{1}{1-\Delta^2 s^2}-\frac{e^{-\Gamma t}/F}{1-(\Delta'^2/F^2)s^2}\right]ds.
\end{equation}

Since both denominators are positive for $s$, the integrand is nonnegative if and only if its numerator 
\begin{equation}
    J(s):=1-\frac{e^{-\Gamma t}}{F}+s^2\left(\frac{e^{-\Gamma t}}{F}\Delta^2-\frac{\Delta'^2}{F^2}\right)
\end{equation}
is not less than 0. Notice that $F>e^{-\Gamma t}$, it remains to prove
\begin{equation}
    E:=e^{-\Gamma t}F\Delta^2-\Delta'^2+F(F-\alpha)\ge 0
\end{equation}
for all admissible parameters.

By introducing $a:=\rho_{11}-\rho_{00}\in[-1,1]$ and $A:=1+e^{-\Gamma t}$, $B:=1-e^{-\Gamma t}$, we have $F=(A-Ba)/2$, $\Delta^2=a^2+4c^2$ and $\Delta'^2=(Aa-B)^2/4+4e^{-\Gamma t}c^2$. For a fixed $a$, $E$ becomes a quadratic function of $c$, with the coefficient of $c^2$ being $-2e^{-\Gamma t}(1+a)\le 0$. Hence $E$ is minimized by maximizing $c^2$, i.e., on the pure state boundary $c^2=(1-a^2)/4$ and $\Delta=1$. Substituting them into $E$ gives
\begin{equation}
\begin{aligned}
    E&\ge\frac{1}{4}\left[(A-Ba)^2-(Aa-B)^2\right]-e^{-\Gamma t}(1-a^2)\\
    &=(1-a^2)\left(\frac{A^2-B^2}{4}-e^{-\Gamma t}\right)\\
    &=0,
\end{aligned}
\end{equation}
therefore $J(s)$ and the integrand in Eq.~\eqref{integral} are nonnegative for all $s\in[0,1]$, thus completing the proof.

\bibliography{./tex/bibspontaneousQRNG}

\end{document}